\newif\ifpreprint
    \preprintfalse        
    \preprinttrue         

\ifpreprint
 	\documentstyle[aas2pp4,epsf,twoside,fleqn]{article}
	\newcommand{\fL}[1]{{\footnotesize\bf\uppercase{#1}}}

	\newcommand{\nc}[1]	{{\,{#1}\,}}
\else
 	\documentstyle[12pt,aasms4,epsf]{article}
	\newcommand{\nc}[1]	{#1}
\fi

\newcommand{\ts}	{\textstyle}
\newcommand{\ds}	{\displaystyle}
\newcommand{\bea}	{\begin{array}}
\newcommand{\eea}	{\end{array}}
\newcommand{\beq}	{\begin{equation}}
\newcommand{\eeq}	{\end{equation}}
\newcommand{\ben}	{\begin{eqnarray}}
\newcommand{\een}	{\end{eqnarray}}
\newcommand{\bsq}	{\begin{mathletters}}
\newcommand{\esq}	{\end{mathletters}}

\newcommand{\eq}	{\nc{=}}	
\newcommand{\mi}	{\nc{-}}	
\newcommand{\pl}	{\nc{+}}	
\newcommand{\id}	{\nc{\equiv}}	
\newcommand{\ap}	{\nc{\approx}}	

\newcommand{\p}		{\partial}
\newcommand{\dx}	{\delta_x}
\newcommand{\dR}	{\delta_R}
\newcommand{\D}		{{\rm d}}
\newcommand{\pr}	{\prime}
\newcommand{\PD}[3]	{\left({\p{#1}\over\p{#2}}\right)_{\!{#3}}}
\newcommand{\JR}	{J_R}
\newcommand{\pR}	{p_R}
\ifpreprint
\newcommand{\tR}	{\theta_{\!R}}
\else
\newcommand{\tR}	{\theta_R}
\fi
\newcommand{\tp}	{\theta_\phi}
\newcommand{\oR}	{\omega_R}
\newcommand{\op}	{\omega_\phi}
\newcommand{\half}	{\case{1}{2}}
\newcommand{\DE}	{\Delta E}
\newcommand{\DL}	{\Delta L^2}
\newcommand{\kmax}	{k_{\rm max}}

\newcommand{\kms}       {\mbox{$\,{\rm km}\,{\rm s}^{-1}$}}
\newcommand{\mkms}	{\mbox{$  {\rm km}\,{\rm s}^{-1}$}}

\newcommand{\eqs}[1]	{equations (\ref{#1})}
\newcommand{\Eqn}[1]	{Equation (\ref{#1})}
\newcommand{\eqn}[1]	{equation (\ref{#1})}
\newcommand{\eqi}[1]	{(equation \ref{#1})}
\newcommand{\eqj}[1]	{equation \ref{#1}}
\newcommand{\eqsj}[1]	{equations \ref{#1}}
\newcommand{\eqb}[1]	{(\ref{#1})}
\newcommand{\Par}[1]	{\S\ref{sec:#1}}
\newcommand{\Sec}[1]	{Section \ref{sec:#1}}

\newcommand{\App}[1]	{Appendix \ref{app:#1}}
\newcommand{\Fig}[1]	{Figure \ref{fig:#1}}

\ifpreprint
  \slugcomment{scheduled for publication in September 1999 issue of 
		The Astronomical Journal}
\else
  \lefthead{Walter Dehnen}
  \righthead{Approximating stellar orbits}
\fi

\begin{document}

\ifpreprint \thispagestyle{empty} \fi
\title{Approximating Stellar Orbits: Improving on Epicycle Theory}
\author{Walter Dehnen\altaffilmark{1}}
\affil{Theoretical Physics, 1 Keble Road, Oxford OX1 3NP, United Kingdom}
\altaffiltext{1}{e-mail: 	dehnen@physics.ox.ac.uk}

\begin{abstract} \ifpreprint\noindent\fi
Already slightly eccentric orbits, such as those occupied by many old stars in 
the Galactic disk, are not well approximated by Lindblad's epicycle theory.
Here, alternative approximations for flat orbits in axisymmetric stellar systems
are derived and compared to results from numeric integrations. All of these
approximations are more accurate than Lindblad's classical theory. I also 
present approximate, but canonical, maps from ordinary phase-space coordinates 
to a set of action-angle variables. 

Unfortunately, the most accurate orbit approximation leads to non-analytical 
$R(t)$. However, from this approximation simple and yet very accurate estimates
can be derived for the peri- and apo-centers, frequencies, and actions 
integrals of galactic orbits, even for high eccentricities. Moreover, further 
approximating this approximation allows for an analytical $R(t)$ and still an 
accurate approximation to galactic orbits, even with high eccentricities.
\end{abstract}


\ifpreprint 	\section{I\fL{ntroduction}} \label{sec:intro} \noindent
\else 		\section{Introduction} \label{sec:intro}
\fi
The knowledge of the stellar orbits is a condition for detailed understanding 
and interpretation of the dynamics of stellar systems such as globular clusters 
and galaxies. Unfortunately, even for fairly simple models for the overall
mass distribution in such systems, the stellar orbits are not accessible in a
closed functional form, and numerical integration of the equations of motion
is necessary. With the progress of computer technology this ceases to be a big 
problem. Nonetheless, it is desirable to be able to approximate stellar orbits 
analytically, in particular it is often useful to have simple and yet accurate 
estimates for the peri- and apo-centers of the stellar orbits and for their 
frequencies and action integrals.

This is often done by first considering a symmetrized version (spherical or 
axisymmetric) of the stellar system in question, and then approximating the 
orbits by perturbation theory. Orbits in axisymmetric galaxies, conserve a 
component $L$ of the angular momentum as well as the energy $E$. Orbits which 
are restricted to the equatorial plane are always regular. Their equation of 
motion can be integrated by first obtaining the radial motion $R(t)$ from
\beq \label{t}
        t  = \int^{R(t)} {\D R^\pr\over\pR} 
           = \int^{R(t)} {\D R^\pr\over 
                        \sqrt{2\big[E-\Phi(R^\pr)\big]-L^2/R^{\pr2}}},
\eeq
and then the azimuthal motion by
\beq \label{phi-t}
        \phi(t) = \int^t \PD{H}{L}{R(t^\pr)}\,\D t^\pr 
		= \int^t {L\,\D t^\pr\over R^2(t^\pr)}
\eeq
where $\Phi(R)$ denotes the gravitational potential, while $H(R,L,p_R)$ is the 
Hamiltonian. Unfortunately, the integral \eqb{t} can be solved analytically 
only for very a few potentials, notably that of a point mass, and the harmonic 
and the isochrone potentials, neither of which gives a good description of the 
potential of real galaxies. Therefore, one often employs an {\em approximate\/} 
solution to \eqb{t}, namely Bertil Lindblad's (1926) epicycle theory, well-known
from the textbooks (cf.\ \cite{bt87}). This approximation may be derived from 
\eqn{t} by expanding the argument of the square root in the denominator into a 
quadratic in $R$, yielding an explicitly soluble integral. However, as we shall 
see below, the argument of the square root is not well approximated by a 
quadratic in $R$, and neglect of the higher-order terms causes the approximation
to be useful only for orbits with quite small eccentricities.

There are two different routes to improve on the classical epicycle motion.
The first one is an extension to higher-order perturbation theory, which can be 
done in various ways (cf.\ \cite{lin58}; \cite{kal79}). The alternative, which 
I will pursue in this paper, is to stay at first-order perturbation theory but 
use a better approximation for the integrand in \eqb{t}. A recipe for doing so
is presented in \Sec{recipe}, while some applications, i.e.\ concrete orbit 
approximations, are given in \Sec{cases}. Lindblad's and Kalnajs' (1979) 
epicycle theories are recovered as the two simplest cases. There are three 
other approximations of interest, all of which are more accurate than these 
epicycle approximations for the important case of a flat rotation curve. 

I also consider the approximate integration of the azimuthal motion \eqb{phi-t} 
with particular emphasis on obtaining a canonical map to the azimuthal angle 
variable $\tp$. This allows for a approximate but canonical map to a set of
action-angle variables $(\JR,L,\tR,\tp)$ and guarantees that the resulting
approximated phase-space flow if incompressible. Such an approximate mapping 
is most simple for the classical epicycle theory (though, to my knowledge, not 
given in the literature).

In \Sec{new}, a new explicit orbit approximation is given, which has been 
derived from the most accurate, but unfortunately implicit, approximation of
\Sec{cases}. As comparisons with numerically integrated orbits in \Sec{comp}
show, this approximation is indeed very accurate, even for highly eccentric
orbits. The paper is summarized and concluded in \Sec{conc}. Appendices A
and B give useful relations for circular orbits and power-law models, 
respectively.

\ifpreprint	\section{A F\fL{ormalism for} A\fL{pproximating} \\
			 S\fL{tellar} O\fL{rbits}} \label{sec:recipe} \noindent
\else		\section{A Formalism for Approximating Stellar Orbits}
		\label{sec:recipe}
\fi
In this section, a general recipe is presented for approximately solving 
\eqn{t} for $R(t)$ and \eqn{phi-t} for $\phi(t)$. I also derive a way for
obtaining an approximate map between ordinary phase-space coordinates and
action-angle variables. Worked examples, i.e.\ concrete orbit approximations
are presented in \Sec{cases}.

\subsection{The Radial Motion} \label{sec:radial}
\subsubsection{Modifying the Time Integral} \label{sec:modify} 
\ifpreprint\noindent\fi
The key idea in obtaining an approximation superior to classical epicycle 
theory is to transform the integral \eqb{t} into a form which is better 
adapted for replacing the argument of the square root by a quadratic.
Consider re-writing \eqn{t} as follows
\beq \label{trans}
	t = \int {\D R\over\pR} 
          = \int {\xi(R)\,\D R\over \xi(R)\,\pR} 
          = \int {x^n\,\D x\over \xi(R)\,\pR},
\eeq
where $\xi(R)$ is a positive definite function of $R$. The relation for $x(R)$ 
follows upon integrating $\xi(R)\,\D R\eq x^n\,\D x$ with integer $n$. We now 
introduce the auxiliary variable $\eta$ by
\beq \label{etat}
	{\D t\over\D\eta} = x^n,
\eeq
which essentially amounts to a new time variable. For $n\nc>0$, \eqn{etat}
means that, for fixed $\D\eta$, the time-step $\D t$ is smaller near peri-center
than near apo-center. With \eqb{etat}, \eqn{trans} becomes
\bsq \label{eta-Y}
\beq \label{eta} 
	\eta = 	  \int {\D x\over\sqrt{Y(x)} }
\eeq with \beq \label{Y}
	Y(x) \equiv 2\,\xi^2(R)\left[E-\Phi(R)-{L^2\over2R^2}\right].
\eeq \esq
The new integral \eqb{eta} is formally identical to the original \eqb{t}, and 
we might derive an approximate solution for $x(\eta)$ in the same way 
Lindblad's epicycle theory is derived from \eqb{t}. Approximating
\beq \label{Y-approx}
	Y \approx X^2 - a^2 (x-x_0)^2,
\eeq
obtained, for example, by Taylor expansion, we find
\beq \label{xeta}
	x \approx x_0 (1-e \cos a\eta),\quad
	e = X / (a\,x_0).
\eeq
With this form of $x(\eta)$, \eqn{etat} is easily solved for
$t(\eta)$. However, $\eta(t)$ cannot be obtained analytically unless $n\eq0$.

\subsubsection{What Functions $\xi(R)$ are Useful?} \label{sec:xi}
\ifpreprint\noindent\fi
Two choices, $\xi\eq1$ and $\xi\eq R$, are of particular interest. For the 
first,
\beq \label{Y-xi-one}
	Y = 2 E - \big[2\Phi(R)+L^2/R^2 \big].
\eeq
For orbits with the same angular momentum $L$, the Taylor expansions of $Y$ are 
identical apart from the additive constant $2E$. The maximum of \eqb{Y-xi-one} 
occurs at the radius $R_L\id R_c(L)$ of the circular orbit with angular 
momentum $L$. At this radius $Y\eq2\DE$ with
\beq \label{DE}
	\DE \equiv E - E_c(R_L),
\eeq
where $E_c(R)$ is the energy of the circular orbit at radius $R$ (see \eqj{Ec-R}
for a definition). 

For the second choice, $\xi\eq R$, 
\beq \label{Y-xi-R}
	Y = 2 R^2 \big[E-\Phi(R)\big] - L^2.
\eeq
This time the Taylor expansions of $Y$ are identical, apart from the additive 
constant ${-}L^2$, for all orbits with the same energy $E$. The maximum of 
\eqb{Y-xi-R} occurs at the radius $R_E\id R_c(E)$ of the circular orbit 
with energy $E$. At this radius
\beq \label{DL}
	Y\eq\DL\equiv L_c^2(R_E) - L^2,
\eeq
where $L_c(R)$ is the angular momentum of the circular orbit at radius $R$ (see
\eqj{Lc-R} for a definition). 

For all other choices of $\xi(R)$, the radius at which $Y$ becomes maximal
depends on both $E$ and $L$ in an essentially non-trivial way.

\subsubsection{The Modified Equation of Motion} \label{sec:eqmot}
\ifpreprint\noindent\fi
We might also give the equation of motion that corresponds to the approximation 
introduced in \Sec{modify}. The integral \eqb{eta} solves
\beq \label{eqxeta}
        {\D^2 x\over\D\eta^2} = {1\over2}\; {\p Y\over\p x},
\eeq
while the approximation \eqb{xeta} actually solves
\beq \label{eqxeta-ap}
        {\D^2 x\over\D\eta^2} = - a^2 (x-x_0)
\eeq
with $X$ being a constant of integration. Note that \eqn{eqxeta} in conjunction 
with \eqn{etat} yields
\beq \label{eqx}
        \ddot{x} = {1\over2}\; {\p (Y\,x^{-2n})\over\p x},
\eeq
which is equivalent to the familiar $\ddot{R}\eq{-}\p H/\p R$. The approximated
equation of motion \eqb{eqxeta-ap} together with \eqn{etat} is equivalent to 
$\ddot{R}\eq{-}\,(\p\hat{H}/\p R)$ with the approximate Hamiltonian
\beq \label{Hhat}
        \hat{H} \equiv E + {\pR^2\over2} - {1\over2\,\xi^2}
                \left[X^2\mi a^2(x\mi x_0)^2\right] + g(E,L).
\eeq
Note that in general $\hat{H}$ depends on $E$ and $L$ as parameters of
$X$, $x_0$, $a$, and, possibly, $x(R)$ and $\xi(R)$. I have added a 
term $g(E,L)$, where $g$ is an arbitrary function, because this addition
does not change the dynamics for $R(t)$.

\ifpreprint	\subsection{The Azimuthal Motion and \\ Action-Angle Variables}
		\label{sec:azi+aav}\noindent
\else		\subsection{The Azimuthal Motion and Action-Angle Variables}
		\label{sec:azi+aav}
\fi
After obtaining an expression for $R(t)$, the radial angle, $\tR$, is easily 
identified as the term, linear in time, of which $R(t)$ is a $2\pi$-periodic
function. With the above formalism, the radial action is
\ifpreprint \ben \else \beq \fi
	J_R 
\ifpreprint &\equiv& \else \equiv \fi
		{1\over2\pi} \oint \pR\,\D R
	     = 	{1\over2\pi} \int_0^{2\pi} 
		\left({\D R\over\D x}\right)^2
		\left({\D x\over\D\eta}\right)^2
		{\D\eta\over\D t}\; \D\eta
\ifpreprint 
	\nonumber \\ &=& 
\else		=
\fi \label{action}
	        {X^2\over2\pi} \int_0^{2\pi}
		{\sin^2\!a\eta\;x^n\,\D\eta \over \xi^2\big(R(x)\big)},
\ifpreprint \een \else \eeq \fi
where $x\eq x(\eta)$ is given in \eqn{xeta}. For this integral to be soluble, 
$\xi(R)$ must not be too complicated when expressed as function of $x$. 

The azimuthal motion may be obtained by inserting the approximation for $R(t)$ 
into \eqn{phi-t} and integrate, if necessary, after further approximating the
integrand. However, the approximate orbits generated in this way do not 
necessarily produce an incompressible flow in phase-space, as Liouville's 
theorem demands. In other words, the part of $\phi(t)$ which is linear in time 
is not, in general, the angle $\tp$ canonical conjugate to the angular momentum 
$L$ at fixed%
\footnote{
	Note that at fixed $(R,\pR)$, the azimuth $\phi$ itself is conjugate 
	to $L$.}
$(\JR,\tR)$. When using such an approximation, spurious effects may result
from this lack. Note that in particular the textbook variant of Lindblad's 
epicycle theory (cf.\ \cite{bt87} p.~124) suffers from this problem; see 
\Par{lb} below for a variant of Lindblad's theory that generates a Hamiltonian
and hence incompressible phase-space flow.

An incompressible phase-space flow or, equivalently, a canonical map to
action-angle variables $(\JR,L,\tR,\tp)$, is best generated from the approximate
Hamiltonian \eqb{Hhat}. That is, the azimuthal motion is approximated 
by integrating $\dot\phi\eq\p\hat{H}/\p L$, and the angle $\tp$ is identified 
as the contribution to $\phi(t)$ which is linear in time.

A complication arises, because the function $g(E,L)$, which appears as additive
term in $\hat{H}$, does affect $\dot\phi$ by adding the constant $\p g/\p L$
to the azimuthal frequency, even though it makes no difference for $R(t)$.
Thus, for the sake of obtaining the relation $\tp(t)$ (but note the canonical
map itself), we must specify the function $g$. This can be done by requiring 
that
\beq \label{dH=or}
        \PD{\hat{H}}{\JR}{L} = \oR,
\eeq
where $\oR$ is the radial frequency emerging from the approximation for $R(t)$.
Inserting the approximation \eqb{xeta} into \eqb{Hhat}, one finds that
$\hat{H}\eq E\pl g$. Together with $\JR$, obtained from \eqn{action}, the 
requirement \eqb{dH=or} gives
\beq \label{dgdE}
	\PD{g}{E}{\!L} = \oR\, \PD{\JR}{E}{\!L} - 1.
\eeq
With this choice of $g(E,L)$, the approximated orbital frequencies satisfy
the identity
\beq \label{op-or}
	\PD{\oR}{L}{\JR}\equiv{\p^2H\over\p\JR\p L}\equiv\PD{\op}{\JR}{L}.
\eeq

\ifpreprint	\section{W\fL{orked} E\fL{xamples}}
\else		\section{Worked Examples}
\fi		\label{sec:cases}
\subsection{Recovering Known Approximations} \label{sec:known}
\placefigure{fig:lb-y}
\subsubsection{Classical Epicycle Theory}
\label{sec:lb}\ifpreprint\noindent\fi
Lindblad's (1926) classical epicycle theory is recovered from the above recipe 
for $\xi\eq1$ and $n\eq0$. The Taylor-expansion of $Y$ around $R_L$ reads, with 
$\dR\eq R\mi R_L$, 
\bsq \label{lb}
\beq \label{lb-y} 
        Y     = 2\DE\mi\kappa^2\,\dR^2
              - \left[{\D\kappa^2\over\D R}{-}{3\kappa^2\over R}\right]_{R_L}
                {\dR^3\over3} + {\cal O}(\dR^4),
\eeq
where the epicycle frequency $\kappa$, defined by \eqn{ka-R}, is evaluated at 
$R_L$, and hence is a function of angular momentum. For the important case of a 
flat rotation curve, \Fig{lb-y} shows $Y\mi2E$ and its approximation by the 
quadratic part of \eqb{lb-y}. The large error of this approximation is related 
to the neglected third-order term in \eqb{lb-y}, which for stellar systems is 
never small, since always $\D\kappa^2/\D R<0$. The radial motion is 
approximated by
\ben 
\label{lb-Rt}	R(t) &=& R_L\,(1-e\cos\tR),				\\
\label{lb-e} 	e    &=& \sqrt{2\,\DE}\,/\,(R_L\,\kappa),		\\
\label{lb-j} 	\JR  &=& \half\,\kappa\,R_L^2\,e^2 = \DE / \kappa,
\een 
where $\tR\eq\kappa t$. The approximate Hamiltonian is, with $E_c\id E_c(R_L)$,
\beq
\label{lb-hH}	\hat{H} = E_c+\half\big(\pR^2+\kappa^2\dR^2\big)
			= E_c+\kappa\,\JR,
\eeq
specifically, $g\eq0$. The azimuthal motion follows from
integrating $\dot{\phi}=\p\hat{H}(R,\pR,L)/\p L$:
\ben
\label{lb-phi}	\phi &=& \tp + e\,\gamma\sin\tR 
			+ e^2\,{\D\kappa\over\D L} R_L^2\sin2\tR	\\
\label{lb-op}	\op  &=& \Omega(L) + {\D\kappa\over \D L} J_R,
\een \esq
where $\gamma\id2\Omega/\kappa$ is evaluated at $R_L$, while $\tp\eq\op t$. It 
appears that these formul\ae\ \eqb{lb-phi} and \eqb{lb-op} for the canonical
map $\phi(\tp)$ and the corresponding azimuthal frequency $\op$ that are 
consistent with the radial motion \eqb{lb-Rt} are not in the literature.
Note that the error of \eqs{lb-phi} and \eqb{lb-op} is still ${\cal O}(e^2)$, 
even though they contain terms of order $e^2\nc\propto\JR$. Omitting those 
terms gives the textbook result for $\phi(t)$, but does not yield a canonical
map to the angle $\tp$ and results in a compressible phase-space flow. The 
azimuthal frequency \eqb{lb-op} and the radial frequency $\oR\eq\kappa(R_L)$, 
satisfy the identity \eqb{op-or}.

\ifpreprint
   \begin{figure}[t]
	\centerline{ \epsfxsize=80mm \epsfbox[20 402 580 710]{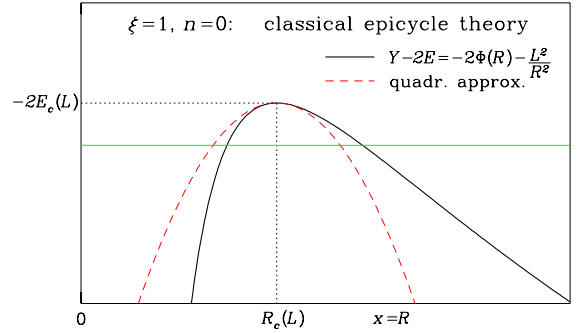}}
	\caption[]{\footnotesize
	Lindblad's classical epicycle theory for the logarithmic potential
	(flat rotation curve). The solid line gives $Y(R)$ \eqb{Y-xi-one}
	while its quadratic approximation \eqb{lb-y} is shown as broken line. 
	The thin horizontal line corresponds to the energy of an orbit with 
	$L\eq0.9L_c(E)$. Note the strong asymmetry of $Y(R)$. \label{fig:lb-y}}
   \end{figure}
\fi

\placefigure{fig:kl-y}
\subsubsection{Kalnajs' Epicycle Theory}
\label{sec:kl}\ifpreprint\noindent\fi
In a little-known four-page paper, Kalnajs (1979) introduced an improved
epicycle theory, which can be derived from our recipe with $\xi\eq\half R$ and 
$n\eq0$, giving $x\eq R^2$. The Taylor expansion of $Y$ \eqi{Y-xi-R} around its 
maximum is, with $\dx\eq R^2\mi R_E^2$,
\bsq \label{kl}
\beq \label{kl-y}
        Y = \DL \mi \frac{1}{4}\kappa^2 \dx^2
            - \left[{\D\kappa^2\over\D R^2}\right]
            {\dx^3\over12} + {\cal O}(\dx^4).
\eeq
Here, $\kappa$ and its derivative are evaluated at $R_E$, i.e.\ they are 
functions of energy. $Y(x)\pl L^2$ and its approximation by the quadratic part 
of \eqb{kl-y} are shown in \Fig{kl-y} for the logarithmic potential. The 
resulting approximation for $R(t)$ is
\ben
\label{kl-Rt}	R(t)&=& R_E\,(1-e\cos\tR)^{1/2},			\\
\label{kl-e}	e   &=& \gamma\,\sqrt{1-L^2/L_c^2},			\\
\label{kl-j}	\JR &=& {L_c\over2\gamma}\left[1-\sqrt{1-e^2}\right],
\een
where $\tR\eq\kappa t$, $\gamma\eq\gamma(R_E)$, and $L_c\id L_c(R_E)$. As 
pointed out by Kalnajs, the neglected third-order coefficient in \eqb{kl-y} is 
always smaller than that in \eqb{lb-y}, implying that his theory is generally 
more accurate than Lindblad's. This is also evident when comparing Figures 
\ref{fig:lb-y} and \ref{fig:kl-y}, where the horizontal lines correspond to 
the same orbit. Kalnajs also remarked that his theory becomes exact for a 
harmonic potential, which arises from a uniform mass distribution and, at the 
time he wrote the paper, was thought to be a good description for the inner 
parts of galaxies.

\ifpreprint
  \begin{figure}[t]
	\centerline{ \epsfxsize=80mm \epsfbox[20 402 580 710]{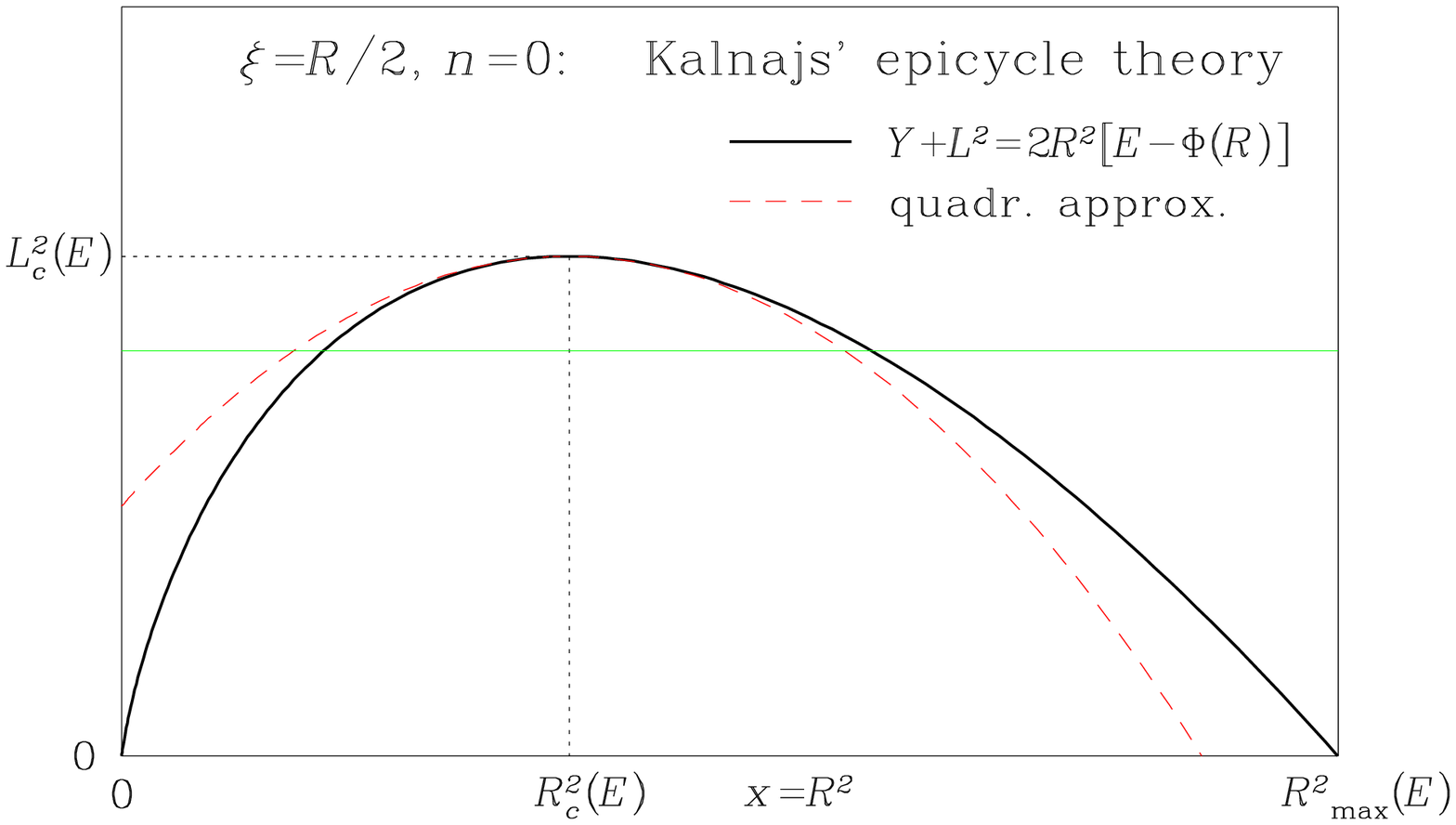} }
	\caption[]{\footnotesize
	Kalnajs' improved epicycle theory for the logarithmic potential.
	The solid line gives $Y(x)$ \eqb{Y-xi-R} while its quadratic 
	approximation \eqb{kl-y} is shown as broken line. The thin horizontal 
	line corresponds to an orbit with $L\eq0.9L_c(E)$. \label{fig:kl-y}}
  \end{figure}
\fi
Integrating $\dot{\phi}\eq\p\hat{H}/\p L$ exactly gives
\ben
\label{kl-phi}	\phi &=& \tp + \sqrt{\gamma^2-e^2\over1-e^2}\; A\!\left(\!
			\sqrt{1+e\over1-e},\,{\tR\over2}\!\right),	\\
\label{kl-op}	\op  &=& \Omega \sqrt{1-e^2\gamma^{-2}\over1-e^2}
			+ {\p g\over\p L},				\\
\label{kl-g} 	g    &=& {1\over2} \int_{E_c(R_L)}^E\!\! \D E\;
                        {\D\ln\kappa^2\over\D\ln R^2}\bigg|_{R_E}
                        \left[1-{1\over\sqrt{1-e^2}}\right]
\een
with $\tp\eq\op t$ and $\Omega\eq\Omega(R_E)$. For the harmonic potential, 
$\Phi\eq\half\omega^2R^2$, $\kappa\eq2\omega$ and $g\eq0$, while for a flat 
rotation curve, $v_c\eq v_0$,
\beq \label{kl-g:flat}
        {\p g\over\p L} = {v_0^2\over2L} \left[1-{1\over\sqrt{1-e^2}}\right].
\eeq\esq
In \eqn{kl-phi}, 
\beq \label{def-A}
        A(x,y) \equiv \arctan\left(x\,\tan y\right) - \arctan(\tan y)
\eeq
is a continuous, periodic function of $y$ with period $\pi$. 

\ifpreprint
  \begin{figure}[t]
	\centerline{ \epsfxsize=80mm \epsfbox[20 402 580 710]{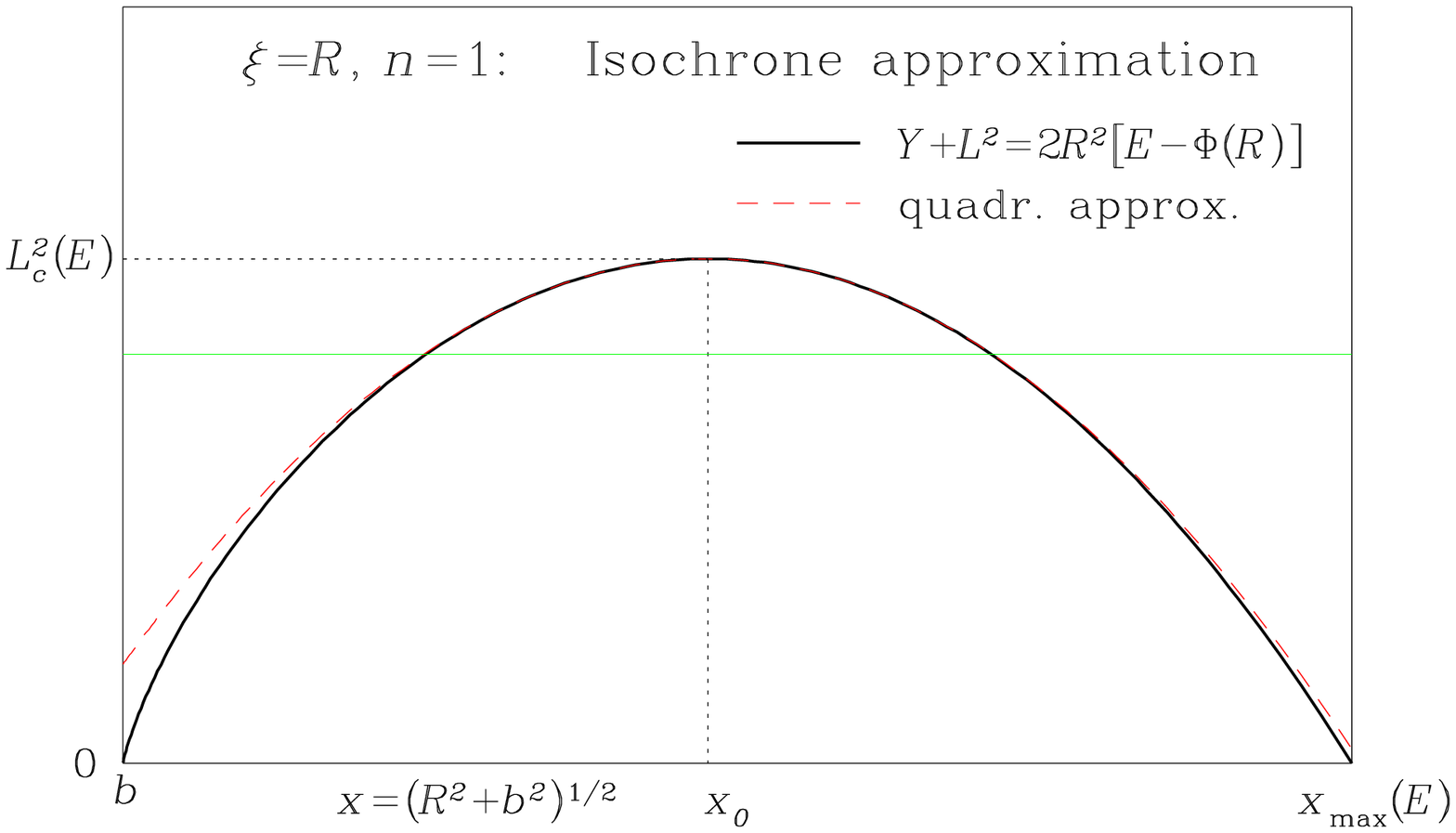} }
	\caption[t]{\footnotesize
	The isochrone approximation for the logarithmic potential. The solid 
	line gives $Y(x)$ \eqb{Y-xi-R} while its quadratic approximation 
	\eqb{iso-y} is shown as broken line. The thin horizontal line 
	corresponds to an orbit with $L\eq0.9L_c(E)$. \label{fig:iso-y}}
  \end{figure}
\fi
\subsection[]{Approximations with \boldmath$\xi\eq R$, $n\eq1$} 
\label{sec:iso+kep}
\placefigure{fig:iso-y}
\subsubsection[]{The Isochrone Approximation} 
\label{sec:iso}\ifpreprint\noindent\fi
For this choice of $\xi$ and $n$, one finds $x\eq \sqrt{R^2\pl b^2}$ where $b$ 
is an arbitrary constant. The Taylor expansion of $Y(x)$ around its maximum at 
$x_0\eq\sqrt{R_E^2\pl b^2}$ is, with $\dx\eq x\mi x_0$ and $\kappa\eq\kappa
(R_E)$,
\beq \label{iso-y}
	Y = \DL \mi x_0^2\,\kappa^2\, \dx^2
	-x_0\!\left[\kappa^2{+}{2x_0^2\over3}{\D\kappa^2\over\D R^2}\right]\!
	\dx^3 + {\cal O}(\dx^4),
\eeq
while the resulting expressions for $R(t)$ and $\phi(t)$ can be found in 
\App{iso}. The neglected third-order term in \eqb{iso-y} can be made to vanish 
identically by the choice
\beq \label{iso-b}
        b^2 = - R_E^2 \bigg({3\over2}
                \bigg[{\D\ln\kappa^2\over\D\ln R^2}\bigg]^{-1}_{R_E}+1\bigg).
\eeq
For all realistic stellar systems, except the harmonic potential, for which 
Kalnajs' theory is exact, \eqn{iso-b} results in well-defined $b^2\nc\ge0$.
\Fig{iso-y} actually plots $Y(x)\pl L^2$ for a flat rotation curve and $b$ 
chosen according to \eqb{iso-b}. A comparison with \Fig{kl-y} shows the 
qualitative improvement over Kalnajs' epicycle theory. This approximation 
becomes exact for the potential for which $b$ is constant, resulting in 
H\'enon's (1959) isochrone sphere
\beq \label{henons}
        \Phi = - {G M\over b+\sqrt{R^2+b^2}}.
\eeq

\placefigure{fig:kep-y}
\subsubsection{The Keplerian Approximation} \label{sec:kep}
\ifpreprint\noindent\fi
I will also consider the simple case $b\id0$. The resulting approximation is 
exact for orbits in the potential of a central point mass, corresponding to 
\eqb{henons} for $b\eq0$. For a flat rotation curve, \Fig{kep-y} plots $Y(x)+
L^2$ and its quadratic approximation. A comparison with \Fig{kl-y} shows that 
for this case the Keplerian approximation is better than Kalnajs' theory. The 
relations for the radial motion are familiar from celestial mechanics (for which
$\gamma\eq2$):
\bsq \label{kp-x} \ben
\label{kp-Rt}	R(t)	&=&	R_E\,(1-e\cos\eta),			\\
\label{kp-tR}	\tR	&=&	\kappa t = \eta - e\sin\eta,		\\
\label{kp-e}	e	&=&	\sqrt{\DL}/(R_E^2\,\kappa)
			 = 	\case{\gamma}{2} \sqrt{1-L^2/L_c^2},	\\
\label{kp-j} 	\JR 	&=& {2L_c\over\gamma}\left[1 - \sqrt{1-e^2}\,\right].
\een
The relation between azimuth and time that results from integrating 
$\dot\phi\eq\p\hat{H}/\p L$ is, with $\tp\eq\op t$, 
\ben
\label{kp-phi} \phi	&=& \tp\pl\sqrt{\gamma^2-4e^2\over1-e^2}\,
		 		\left[{e\sin\eta\over2}+
				A\!\left(\!\sqrt{1+e\over1-e},
				\,{\eta\over2}\!\right)\right],		\\
\label{kp-op}	\op	&=& \Omega\,\sqrt{1-4e^2\gamma^{-2}\over1-e^2} 
				+ {\p g\over\p L},			\\
\label{kp-g}	g	&=& \int_{E_c(R_L)}^E\!\!\! \D E\; 
				\bigg[3+2{\D\ln\kappa^2\over\D\ln R^2}
				\bigg|_{R_E}\bigg]\,\bigg[1-
				{1\over\sqrt{1-e^2}}\bigg].
\een
For the potential generated by a central point mass, $g\eq0$, while for a flat 
rotation curve,
\beq
        {\p g\over\p L} = -{v_0^2\over L} \left[1-{1\over\sqrt{1-e^2}}\right].
\eeq \esq

\ifpreprint
  \begin{figure}[t]
	\centerline{ \epsfxsize=80mm \epsfbox[20 402 580 710]{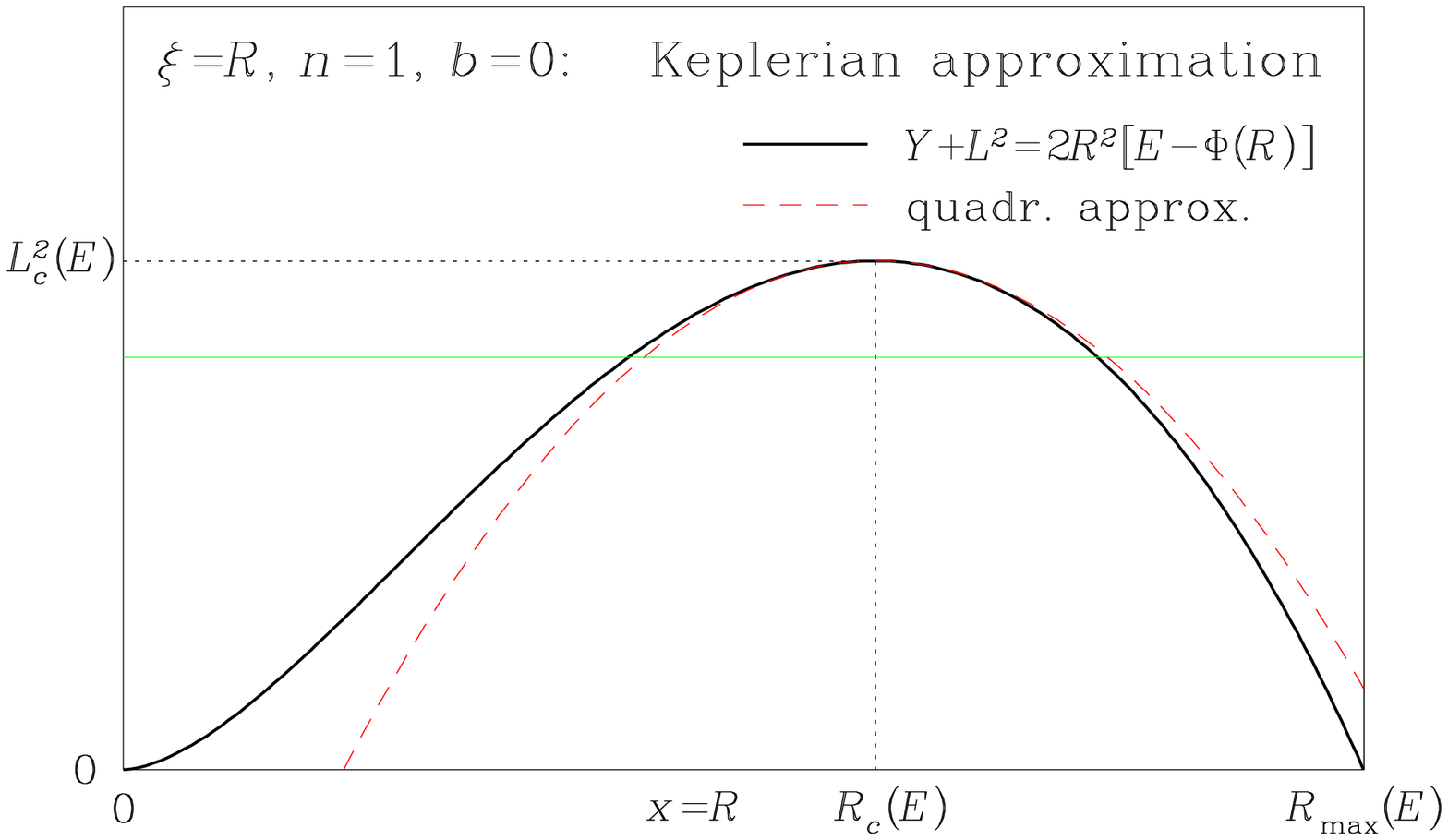} }
	\caption[]{\footnotesize
	The Keplerian approximation for the logarithmic potential. The solid 
	line gives $Y(x)$ \eqb{Y-xi-R} while its quadratic approximation 
	\eqb{iso-y} is shown as broken line. The thin horizontal line 
	corresponds to an orbit with $L\eq0.9L_c(E)$. \label{fig:kep-y} }
  \end{figure}
\fi
\placefigure{fig:crz-y} 
\subsection[]{\boldmath$\xi\nc\propto R$ and $n\eq\gamma(R_E)\mi1$} 
\label{sec:crz} \ifpreprint\noindent\fi
The approximation that results from our recipe for 
this choice of $\xi$ and $n$, giving
\bsq \label{crz} \beq \label{crz-x}
	x = (R/R_E)^{2/\gamma}, 
\eeq
is of particular interest. For the harmonic and Kepler potential, this yields, 
respectively, Kalnajs' epicycle theory and the Keplerian approximation, i.e.\ 
the exact solutions. For other potentials, $n$ is non-integer, and consequently 
$t(\eta)$ is not analytical. However, one may still give $R$ and $\pR$ as 
functions of $\eta$ and evaluate $\JR$. Here are the basic results (with 
$\dx\eq x\mi1$ and $L_c\eq L_c(R_E)$)
\ben
\label{crz-y}	Y	&=& L_c^2\left[e^2\mi\dx^2
			    	\mi\left(\gamma{-}1{+}\case{\gamma}{3}
				\case{\D\ln\kappa^2}{\D\ln R^2}\right)\dx^3	
				\pl{\cal O}(\dx^4)\right],		\\
\label{crz-e}	e	&=& \sqrt{1-L^2/L_c^2},				\\
\label{crz-Rt}	R	&=& R_E\, \big(1-e\cos\eta\big)^{\gamma/2},	\\
\label{crz-t}	\kappa t&=& {\ts \int}_0^\eta \big(1-e\cos\eta^\pr
				\big)^{\gamma-1}\,\D\eta^\pr,		\\
\label{crz-j}	\JR	&=& \case{\gamma}{2}(L_c-|L|).
\een
Note that the neglected third-order term in \eqb{crz-y} is usually small%
\footnote{
	Replacing \eqb{crz-x} with $x^\gamma\eq (R^2+b^2)/R_E^2$ and choosing
	the constant $b$ such as to set the third-order term to
	zero, as in the isochrone approximation, does not work, since it 
	requires $b^2\nc<0$.}:
in the case of power-law rotation curves, $v_c\nc\propto R^\beta$, its 
coefficient becomes maximal in magnitude (at $-0.0572\dots$) for $\beta\eq
0$. For this worst case of a flat rotation curve, \Fig{crz-y} plots $Y(x)$ and 
its quadratic approximation. Evidently, this approximation is superb, even for 
small angular momenta. In particular, it always gives a peri-center radius of 
zero for $L\eq0$, which all the other approximations fail to do (apart from the 
cases where they are exact). Note that the radial frequency 
\beq \label{crz-oR}
	\oR = \kappa\,\left[\big(|L|/L_c\big)^{\gamma-1}\,
			P_{\gamma-1}\big(L_c/|L|\big)\right]^{-1},
\eeq
where $P_\nu(x)$ denotes the associated Legendre function,
is larger than $\kappa$, unless $L\eq L_c$, $\gamma\eq1$, or $\gamma\eq2$.

Integrating $\dot{\phi}\eq \p\hat{H}/\p L$ gives
\ben 
\label{crz-phi} \phi &=&  \tp + \gamma\left[{\eta-\tR\over2}+A\left(
			\sqrt{1+e\over1-e},{\eta\over2}\right)\right],	\\
\label{crz-op}	\op  &=&  \Omega\,{\oR\over\kappa} + {\p g\over\p L},	\\
\label{crz-g}	g    &=& \int_{E_c(R_L)}^{E}\!\!\!\D E\left(
		{\oR\over\kappa}\left[1-\gamma^2{\ts{\D\ln\gamma\over\D\ln R^2}}
		\Big(1\mi{\ts{|L|\over L_c}}\Big)\right]\mi1\right).
\een
with $\tp\eq\op t$ and $\tR\eq\oR t$. For a flat rotation curve,
\beq \label{crz-gl}
	{\p g\over\p L} = -{\Omega\over\sqrt{1-e^2}}
		\left[{\oR\over\kappa}-1\right].
\eeq \esq

\ifpreprint
  \begin{figure}[t]
	\centerline{ \epsfxsize=80mm \epsfbox[20 402 580 710]{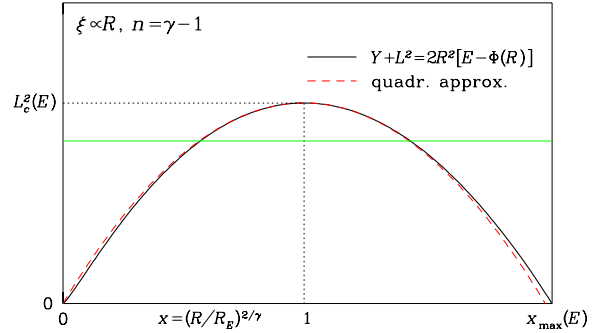} }
	\caption[t]{\footnotesize
	The approximation arising from $\xi\nc\propto R$ and $n\eq\gamma\mi1$ 
	for the logarithmic potential. The solid line gives $Y(x)$ \eqb{Y-xi-R},
	while its quadratic approximation \eqb{crz-y} is shown as broken line.
	The thin horizontal line corresponds to an orbit with $L\eq0.9L_c(E)$.
	\label{fig:crz-y} }
  \end{figure}
\fi
\ifpreprint
  \begin{figure}[t]
	\centerline{ \epsfxsize=80mm \epsfbox[20 402 580 710]{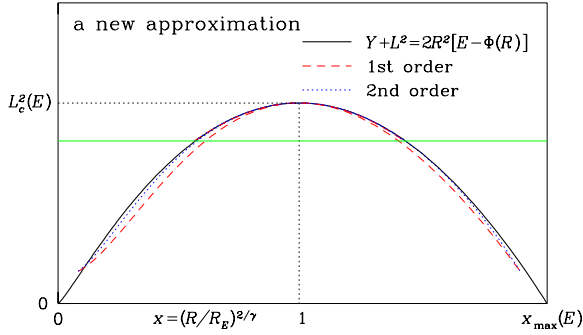} }
	\caption[t]{\footnotesize
	The approximations described in \Par{new} for the logarithmic potential.
	The solid line gives $Y(x)$ \eqb{Y-xi-R}, while the dashed and dotted 
	lines correspond to the approximation \eqb{new-y} for $\kmax\eq1$ and
	$\kmax\eq2$, respectively, for the two orbits with $L/L_c(E)\eq0.4$ and
	0.9 (for the latter the approximations can hardly be distinguished from
	the exact relation).
	\label{fig:new-y} }
  \end{figure}
\fi
\placefigure{fig:new-y} 
\ifpreprint	\section{A N\fL{ew} O\fL{rbit} A\fL{pproximation}} 
		\label{sec:new}\noindent
\else		\section{A New Orbit Approximation} \label{sec:new}
\fi
The remarkably accurate orbit approximation presented in \Par{crz} above has
the unfortunate drawback, that the time $t$ cannot explicitly solved for,
because the integral in \eqn{crz-t} is generally not analytically soluble. 
However, one may approximately solve this integral, by replacing its integrand 
by a finite powers series in $e$. If the highest power retained in this 
procedure is $e^{\kmax}$, this technique amounts to approximating, in the 
formalism of \Par{recipe}, 
\bsq \label{new}
\beq \label{new-y}
	Y = R^2p_R^2 \approx  L_c^2\,(e^2-\dx^2)\;\left[{x^{\gamma-1}\over
	\sum_{k=0}^{\kmax} {\gamma-1\choose k} \dx^k} \right]^2.
\eeq
For the case of a flat rotation curve, \Fig{new-y} plots $Y(x)$ and the 
approximation \eqb{new-y} with $\kmax\eq1$ (dashed) and $\kmax\eq2$ (dotted) for
the two orbits with $L/L_c\eq 0.4$ and 0.9. Evidently, these approximations 
are not much worse than their parent, i.e.\ \eqn{crz-y}, and at least as good 
as the isochrone approximation above. The radial motion, $R(t)$, is given by
\ifpreprint \ben
\label{new-Rt}	R	&=& R_E\, \big(1-e\cos\eta\big)^{\gamma/2},	\\
\label{new-e}	e	&=& \sqrt{1-L^2/L_c^2},				\\
		\kappa t&=& \eta - e\,(\gamma{-}1)\sin\eta	
		\nonumber \\ & & \label{new-t} 
				+ \case{1}{4}e^2(\gamma\mi1)(\gamma\mi2)
				(\eta+\half\sin2\eta) + \dots,		\\
\label{new-oR} \oR &=& \kappa\,\left[1+\case{1}{4}e^2(\gamma\mi1)
			(\gamma\mi2)+\dots\right]^{-1}.
\een \else \ben
\label{new-Rt}	R	&=& R_E\, \big(1-e\cos\eta\big)^{\gamma/2},	\\
\label{new-e}	e	&=& \sqrt{1-L^2/L_c^2},				\\
\label{new-t}	\kappa t&=& \eta - e\,(\gamma{-}1)\sin\eta	
				+ \case{1}{4}e^2(\gamma\mi1)(\gamma\mi2)
				(\eta+\half\sin2\eta) + \dots,		\\
\label{new-oR} \oR &=& \kappa\,\left[1+\case{1}{4}e^2(\gamma\mi1)
			(\gamma\mi2)+\dots\right]^{-1}.
\een \fi
\Eqn{new-oR} implies $\oR\nc\ge\kappa$ since always $1\nc\le\gamma\nc\le2$, and
is, for power-law rotation curves, even more accurate than its parent 
\eqb{crz-oR}, i.e.\ including terms ${\cal O}(e^4)$ would degrade the accuracy 
of the new approximation. Note that the corresponding radial action cannot be 
given in closed form. The azimuthal motion may be approximated using 
\eqn{crz-phi} and replacing $\p g/\p L$ by its result for a flat rotation 
curve (which for power-law models makes an error of the order $e^4$):
\ben
\label{new-phi} \phi &=& \op t + \gamma \left[{\eta-\oR t\over2} + 
		A\left(\sqrt{1+e\over1-e},{\eta\over2}\right)\right], \\
\label{new-op} \op  &=& \Omega\left[{\oR\over\kappa}-{1\over\sqrt{1-e^2}}
		\Big({\oR\over\kappa}-1\Big)\right].
\een \esq
The relation \eqb{new-t} for $t(\eta)$ is generally not invertible. As a 
consequence, one must either (i) use $\eta$ rather than $t$ as independent 
variable, (ii) solve \eqb{new-t} numerically for $\eta(t)$, or (iii) use the 
approximate inversion
\ifpreprint \ben \else \beq \fi
	\eta 
\ifpreprint &\approx& \else \approx \fi
	\oR t + e(\gamma\mi1) \sin\oR t
\ifpreprint \nonumber \\ & & \fi \label{new-eta}
	+ \case{1}{8}e^2(\gamma\mi1)(3\gamma\mi2) \sin2\oR t.
\ifpreprint \een \else \eeq \fi
Note, however, that using this relation is {\em not\/} equivalent to using
\eqb{new-t}, rather it generates a different orbit approximation.

As we will see below, the orbit approximation defined by \eqs{new} is remarkably
accurate down to high eccentricities. However, it is not Hamiltonian, i.e.\ the 
phase-space flow generated is in general not incompressible. It deviates from 
this ideal by an amount of the order $e^{\kmax+1}$.

\ifpreprint
  \begin{figure*}
	\centerline{ \epsfxsize=120mm \epsfbox[18 324 589 707]{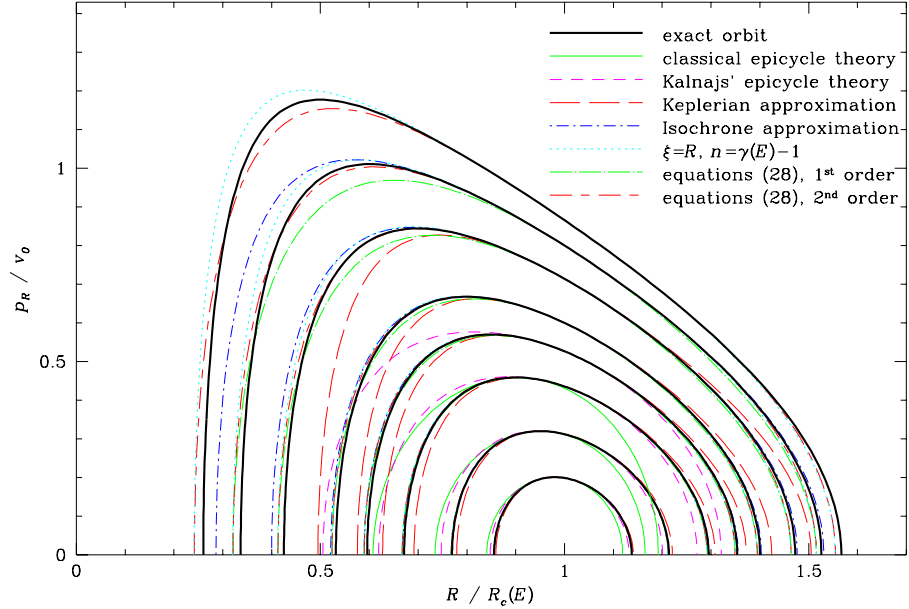}}
	\caption[]{\footnotesize
	Surfaces of the $R$-$\pR$ section of phase space: comparison of exact 
	orbits in the logarithmic potential, which supports a flat rotation 
	curve, with the predictions of the various orbit approximation of 
	\Par{cases} and \Par{new}. The orbital angular momenta are, from inside
	out, 0.98, 0.95, 0.9, 0.85, 0.8, 0.7, 0.6, and 0.5 of $L_c(E)$.
	\label{fig:RpR} }
  \end{figure*}
\fi
\ifpreprint
  \begin{figure*}
	\centerline{ \epsfxsize=120mm \epsfbox[18 324 589 707]{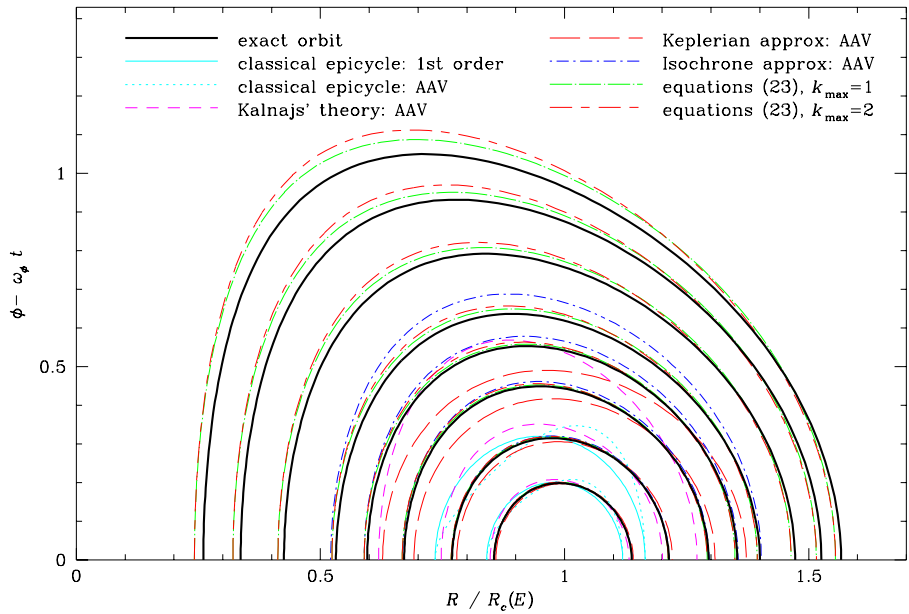}}
	\caption[]{\footnotesize
	Surfaces of the $R$-$(\phi\mi\op t)$ section of phase space: 
	comparison of exact orbits in the logarithmic potential, which supports 
	a flat rotation curve, with the approximations for the azimuthal motion.
	The orbits have the same angular momenta as in \Fig{RpR}. Approximations
	which allow for a canonical map to a set of action-angle variables are 
	indicated as `AAV'. Note that multiplying the $y$-axis with $R$, these 
	essentially are the epicycles which the orbits perform around their 
	guiding centers.
	\label{fig:Rphi} }
  \end{figure*}
\fi

\ifpreprint	\section{C\fL{omparison with} E\fL{xact} O\fL{rbits}}
		\label{sec:comp}\noindent
\else		\section{Comparison with Exact Orbits}\label{sec:comp}
\fi		
We now compare the various orbit approximations with numerically integrated
orbits in the logarithmic potential, $\Phi\eq v_0^2\ln(R/R_0)$, which supports
a flat rotation curve. This potential is scale invariant, i.e.\ all orbits with
the same ratio $|L|/L_c(E)$ but different energies are identical, apart from 
a scaling relation or the sense of rotation.

\placefigure{fig:RpR}

\subsection{Testing the Radial Motion} \label{sec:comp:rad}
\ifpreprint\noindent\fi
For eight orbits with $|L|/L_c(E)$ between 0.5 and 1, \Fig{RpR} compares the 
plots of $R$ vs.\ $\pR$ (surfaces of section) predicted from the various orbit 
approximations with the exact relations. The agreement in this plot is 
indicative of the accuracy of the relation for $R(t)$. For a quantitative 
comparison with the orbit of a star with velocity $(U,V,W)$ in the solar 
neighborhood, one should compare $\pR$ with the radial velocity $U$. The 
azimuthal velocity $V$ w.r.t.\ the local standard of rest can be evaluated
from the ratios $R/R_c(E)$ and $L/L_c(E)$ to be $V\eq\big([L/L_c(E)]/[R/R_c(E)
]-1\big)v_0$ with $v_0$ denoting the local circular speed.

Evidently, Lindblad's classical epicycle theory becomes significantly inaccurate
for $|L|\nc\la0.98 L_c$, corresponding in the solar neighborhood to stars 
with $|U|\ga45$\kms\ or $|V|\ga32$ \mkms\ (with $v_0\ap220\kms$). A 
significant fraction of old-disk stars in the solar neighborhood have velocities
in excess of these values (cf.\ Dehnen, 1998), i.e.\ the classical epicycle 
theory cannot be safely used in quantitative studies of the local Galactic disk.

Kalnajs' epicycle theory is applicable down to $|L|\approx0.95L_c(E)$. 
Theories with $n\eq1$ are better than the epicycle theories ($n\eq0$). 
The Keplerian approximation predicts too large peri-centers for $|L|\nc\la0.85
L_c(E)$, and is thus just capable of describing $R(t)$ for stars in the 
high-velocity tail of the local stellar disk distribution. The isochrone 
approximation is superior to these, giving accurate approximations down to 
$|L|\ap0.7L_c(E)$. 

Also shown are the relations arising from the approximations in \Par{crz} and
\Par{new}. Evidently, these are very good: the approximation derived in 
\Par{crz} and its second-order expansion (\eqsj{new} with $\kmax\eq2$) are much 
better than any of the other approximations. Even the first-order expansion 
($\kmax\eq1$) is comparable in accuracy to the isochrone approximation, which 
itself is second-order accurate.

\ifpreprint
  \begin{figure*}
	\centerline{ \epsfxsize=140mm \epsfbox[21 290 586 710]{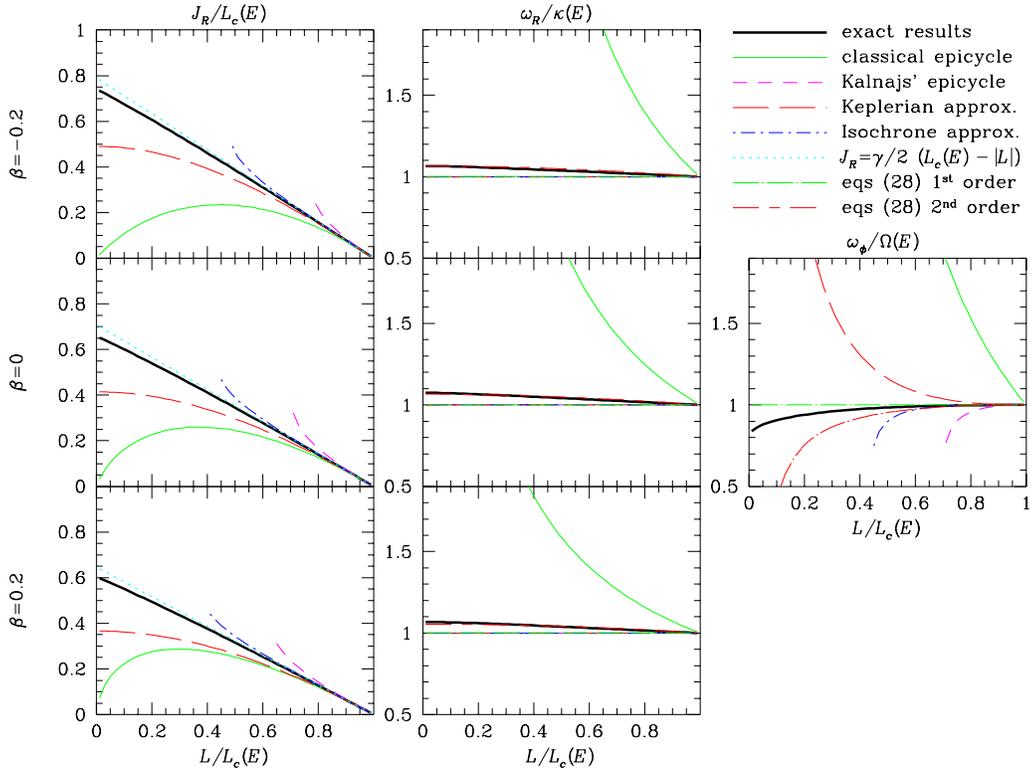}}
	\caption[]{\footnotesize
	Radial action $\JR$ (left column) and frequency $\oR$ (middle column)
	and azimuthal frequency $\op$ (right column): comparison with exact 
	orbits for the potentials supporting the rotation curves $v_c\nc\propto 
	R^\beta$ with $\beta\eq{-}0.2$, 0, and 0.2 (from top to bottom). Note
	that $\oR\eq\kappa(E)$ for all these approximations except Lindblad's 
	epicycle theory and the approximation of \Par{new} with $\kmax\eq2$
	(\eqj{new-oR}, second order). \label{fig:joo} }
  \end{figure*}
\fi

\placefigure{fig:Rphi}
\subsection{Testing the Azimuthal Motion} \label{sec:comp:azi}
\ifpreprint\noindent\fi
To test the quality of the approximations for the azimuthal motion, I compare 
in \Fig{Rphi} the surfaces of $R$-($\phi\mi\op t)$ sections of the same
orbits as in \Fig{RpR} with some of the approximtions. 

The approximations for which $\phi(t)$ has been obtained such as to yield a 
canonical map to a set of action-angle variables, are indicated as `AAV' in the
figure index. Among these approximation, the order of accuracy is similar as 
for the radial motion, but they are less accurate in this section of phase 
space than in the $R$-$\pR$ section. This is not surprising, as the 
approximations were designed to approximate $R(t)$, and any deviation from the 
true $R(t)$ is likely to be amplified when integrating for $\phi(t)$.

\Fig{Rphi} also plots three orbit approximations which do not yield a canonical
map to action-angle variables: the textbook variant of the classical epicycle 
theory and the new approximation of \Par{new} for first- and second-order 
expansions. This variant of the epicycle theory is clearly superior in accuracy
to its canonical counterpart, i.e.\ enforcing a canonical mapping degrades the 
accuracy of the approximation. The two approximations resulting from \eqs{new} 
are much more accurate than all the others, though they are slightly less 
accurate than in the $R$-$p_R$ surface of section. Strangly, the first-order 
approximation is more accurate than the second-order.

\placefigure{fig:joo}
\subsection{Testing Frequencies and the Radial Action} \label{sec:comp:joo}
\ifpreprint\noindent\fi
In \Fig{joo}, the radial action and orbital frequencies predicted by the orbit 
approximations are compared with the exact ones for all orbits in
the potentials supporting nearly flat rotation curves $v_c\nc\propto R^\beta$ 
with $\beta\eq{-}0.2$, 0, and 0.2. Even for nearly eccentric orbits, the 
classical epicycle approximation under-estimates $\JR$ and drastically 
over-estimates the frequencies. The three approximations with $\xi\nc\propto R$
and integer $n$ (Kalnajs, Keplerian, isochrone) have $\oR\eq\kappa(R_E)$,
which actually is not a bad description of the truth. Similarly, $\op\eq
\Omega(R_E)$ is a reasonably good estimate. The estimates for $\oR$ from the 
orbit approximations with canonical maps to action-angle variables are only 
useful for large $L$, corresponding to small $e$. 

I also plotted the estimates \eqb{new-oR} and \eqb{new-op} for $\oR$ and $\op$ 
respectively. While the latter becomes inaccurate at $L\nc\la0.5L_c$, the
estimate for $\oR$ is very accurate for all $L$ (the corresponding lines are
almost entirely overlaid by those for the exact relations).

At $L\ap L_c(E)$, all orbit approximations give $\JR$ correctly, insomuch that 
they agree in the linear term when expanding $\JR$ as power series in $1\mi|L|
/L_c$.  This linear relation is actually the result for $\JR$ obtained from the 
approximation of \Par{crz}.

\ifpreprint	\section{C\fL{onclusions}}	\label{sec:conc}\noindent
\else		\section{Conclusions}		\label{sec:conc}
\fi		
The classical epicycle theory of Lindblad (1926) is usually derived starting 
from the Hamiltonian or, equivalently, from the equations of motions for a 
near-circular orbit in a spherical or flat axisymmetric potential. In this 
paper, I instead directly considered the integral \eqb{t} to be solved in 
order to obtain $R(t)$. The simplest way to approximate this integral by some 
closed form leads straight to the classical epicycle theory. In \Sec{recipe}, 
I derive a general method for approximating this integral in a better way by 
manipulating the integrand before approximating it. 

\subsection{Relation to Other Work} \label{sec:shu}\ifpreprint\noindent\fi
Several published studies deal with improved approximations for the stellar 
orbits. Most of them start with Lindblad's classical epicycle theory and try to
improve it by extension to higher orders, variations of the constants involved, 
or both. Apart from the quoted paper of Kalnajs, I know of only one more study, 
Shu (1969), in which an orbit approximation is derived that does not incorporate
Lindblad's theory. 

Shu also considers directly the integral \eqb{t} and manipulates it to obtain 
approximate solutions. Unlike this work, Shu uses a (truncated) power-series 
expansion in the eccentricity, the lowest order of which yields Lindblad's 
theory, while the next order yields an approximation that becomes exact in the 
case of Keplerian motion. This approximation is not quite the same, though, as 
the Keplerian approximation above, for example, the radial frequency for Shu's 
theory is $\oR\eq(1\mi e^2)^{3/2}\kappa(R_L)$ with $e$ given in \eqn{lb-e},
while the approximation proposed in \Sec{kep} has $\oR\eq\kappa(R_E)$.

\subsection{New Orbit Approximations} \ifpreprint\noindent\fi
From the general recipe in \Sec{recipe}, endlessly many orbit approximations 
may be derived. In \Sec{cases}, four of them are presented in addition to the 
classical epicycle theory. One has already been introduced by Kalnajs (1979) 
and becomes exact for the harmonic potential. Two other approximation yield the 
exact orbits for the potentials of, respectively, a central point mass and 
H\'enon's (1959) isochrone sphere, and I call them accordingly Keplerian and 
isochrone approximations. A comparison with exact orbits in the logarithmic 
potential, which supports a flat rotation curve, shows that the latter two 
approximations are better than Kalnajs', which in turn is superior to the 
classical epicycle theory. In particular, they may safely be applied even to 
the high-velocity tail of the stellar old-disk population.

The fourth new orbit approximation may be considered a hybrid between Kalnajs' 
theory and the Keplerian approximation, it is very accurate, even for purely 
radial orbits, for potentials with power-law rotation curves, $v_c\nc\propto 
R^\beta$. However, this approximation has the severe drawback that the time
cannot explicitly be solved for.

\subsection{Action-Angle Variables} \ifpreprint\noindent\fi
After obtaining an approximation for $R(t)$, I also consider the task of 
integrating the azimuthal motion such that the resulting approximation for the 
phase-space flow is incompressible, as Liouville's theorem demands. This is 
equivalent to obtaining an approximate but canonical map to a set of 
action-angle variables $(\JR,L,\tR,\tp)$. For Lindblad's and Kalnajs' epicycle 
theories, as well as for the new Keplerian and isochrone approximation, this 
map can be given explicitly. However, the azimuthal frequency $\op$ that is 
consistent with this map cannot be expressed in closed form, except for the 
classical epicycle theory or otherwise for special potentials (the logarithmic 
potential and that for which the respective approximation is exact). 

Using an approximation for the azimuthal motion that does not yield an
incompressible phase-space flow may lead to spurious effects. An example
for such an approximation is the textbood variant of Lindblad's classical
epicycle theory. In \Sec{lb}, a version of this theory is presented that does
not suffer from this problem. 
 
\subsection{Estimates of Orbital Properties} \ifpreprint\noindent\fi
In face of the ease by which stellar orbits can be computed numerically, orbit
approximations may be most important as a tool for quantitative estimates of 
stellar dynamical processes, in particular in dynamically cool stellar disks. 
However, a comparison of the approximations with integrated orbits in \Sec{comp}
showed that the classical epicycle theory greatly over-estimates the orbital 
frequencies and under-estimates the peri- and apo-center and radial action. 
Consequently, quantitative analyses based on this approximation will {\rm 
inevitably\/} involve systematic errors. The size of these error may reach 10\%
of more already for slightly eccentric orbits, which in the solar neighborhood 
correspond to velocities w.r.t.\ LSR in excess of about 40\kms.

Highly accurate estimates for the orbital quantities may be derived from the 
orbit approximation presented in \Sec{crz}. This approximation, even though it
does not yield $R(t)$ in closed form, is very accurate for orbits in potentials 
which support nearly flat rotation curves, like that of the Galaxy. The
resulting estimates for the peri- and apo-centers are
\bsq \label{est} \beq \label{Rpm}
	R_\pm(E,L) \approx R_c(E)\,(1\pm e)^{\gamma(E)/2}
\eeq with eccentricity \beq \label{ecc}
	e^2	= 1-L^2/L_c^2(E)
\eeq
and $\gamma\id2\Omega/\kappa$. Here, $R_c(E)$, $L_c(E)$, $\Omega(E)$, and 
$\kappa(E)$ denote the radius, angular momentum, azithmal frequency, and 
epicyclic frequency, respectively, of the circular orbit with the same energy 
$E$ as the orbit in question. For orbits in potentials with power-law rotation 
curves $v_c\nc\propto R^\beta$, \eqn{Rpm} under-estimates the apo-center by at 
most 1.7\% (for $L\eq0$, $\beta\ap{-}0.16$), while the peri-center is 
under-estimated by at most $\nc\sim2\%$ (for $L\ap0.3L_c$, $\beta\ap0.1$). The 
corresponding estimate for the radial action is
\beq \label{JR}	
	\JR(E,L) \approx \half\,\gamma(E)\,\big(L_c(E)-|L|\big).
\eeq
For power-law rotation curves, the largest error of this estimate occurs on
purely radial orbits ($L\eq0$) in the logarithmic potential ($\beta\eq0$),
where actually $\JR\eq L_c\sqrt{{\rm e}/2\pi}$ on radial orbits \eqi{JoLc}, 
while the above approximation predicts a value 7.5\% larger. 

Another way to use this very accurate orbit approximation is to further 
approximate it, in order to overcome its major drawback and solve for the
time explicitly. This results in a new orbit approximation, presented in 
\Sec{new}, which gives $R(t)$ and is still very accurate to high 
eccentricities. The corresponding approximation for the radial frequency
\beq \label{or}
	\oR(E,L) \approx \kappa(E) / \big(1+\case{e^2}{4}(\gamma-1)(\gamma-2)
		\big)
\eeq
is, for power-law rotation curves, accurate to better than 1.4\% (at $L\eq0$, 
$\beta\ap0.3$), while the rough estimate $\oR(E,L)\approx\kappa(E)$ 
under-estimates the radial frequency by at most 7.5\% (at $L\eq0\eq\beta$). The
azimuthal frequency, which may be expressed as the time average of $LR^{-2}$, 
is the orbital quantity most difficult to estimate, because it is very sensitive
to $R(t)$ near peri-center, where the approximations for the radial motion are 
least accurate. Numerical experiments suggest that the simple estimate
\beq \label{op}
	\op \approx \Omega(E)
\eeq \esq
is still the most useful: it becomes inaccurate at the ${\ga}10\%$ level for 
orbits with $L\nc\la0.3L_c$.

Equations \eqb{Rpm} to \eqb{op} provide accurate and yet simple estimates for 
basic orbital properties in typical galactic potentials. Note however, that 
these estimates are not self-consistent, i.e.\ they do not in general satisfy 
$\oR\eq(\p E/\p\JR)_L$ or $\op\eq(\p E/\p L)_{\JR}$.

\acknowledgements \ifpreprint\noindent\fi
It is a pleasure to thank James Binney for many stimulating discussions and
reading an early version of this paper. I also thank the referee Agris Kalnajs
for careful comments. This work was financially supported by PPARC.

%
\appendix
%

\ifpreprint	\section{C\fL{ircular} O\fL{rbits}}
		\label{app:circ}\noindent
\else 		\section{Circular Orbits}
		\label{app:circ}
\fi
This appendix summarizes the relations between the quantities describing 
circular orbits. While some of these relations are well-known, several others,
even though trivial in principle, are less familiar. A circular orbit
at radius $R$ is defined by equilibrium between gravitation and centrifugal
forces, resulting in the relation 
\beq \label{vc}
	v_c^2(R)\eq R^{-1}(\p\Phi/\p R)
\eeq
for its azimuthal velocity. A circular orbit is uniquely determined by either 
its radius $R$, angular momentum $L$, energy $E$, circular frequency $\Omega$, 
or radial frequency $\kappa$, also known as {\em epicycle\/} frequency. (Radial
oscillations are not excited for circular orbits, but their radial frequency is 
well-defined nonetheless). The definitions and some relations between these are
\ben 
\label{Lc-R}	L_c^2	&\equiv& R^2\,v_c^2
			 = R^3 {\p\,\Phi\over\p R} 
			 = R^4\,\Omega^2;	\\
\label{Ec-R} 	E_c	&\equiv& {v_c^2\over2} + \Phi
			 ={\p(R^2\,\Phi)\over\p R^2};			\\
\label{Om-R}	\Omega^2&\equiv& R^{-2}\, v_c^2
			 = R^{-4}\,L_c^2 
			 = R^{-1} {\p\,\Phi\over\p R}
			 = \left({\D E_c\over\D L_c}\right)^2;		\\
\label{ka-R}	\kappa^2&\equiv&\left({\p^2\over\p R^2}\!\Big[\Phi
			   	\pl{L^2\over2R^2}\Big]\right)_{\!L=L_c}
			 = {\p^2\Phi\over\p R^2}
			   	+{3\over R}{\p\,\Phi\over\p R}		\\
\label{ka-a}		&=& {1\over R^3}{\D L^2_c\over\D R}
			 =  R {\D\,\Omega^2\over\D R} + 4 \,\Omega^2	\\[1ex]
\label{ka-b}		&=& 4{\D E_c\over\D R^2}
			 =  4{\p^2(R^2\,\Phi)\over\p(R^2)^2};		\\
\label{dOm}	{\D\Omega\over\D E} &=& {1-\gamma^2\over L_c},
\een
where
\beq
	\gamma \equiv {2\Omega\over\kappa}.
\eeq

\ifpreprint	\section{P\fL{ower}-L\fL{aw} P\fL{otentials}}
		\label{app:powlaw}\noindent
\else 		\section{Power-Law Potentials}
		\label{app:powlaw}
\fi
Quite often one is dealing with simple power-law models with circular speed
\beq\label{pl-vc}
        v_c(R) = v_0 \left({R\over R_0}\right)^\beta
\eeq
and gravitational potential
\beq\label{Phi}
        \Phi(R) = v_0^2 \times \left\{ \bea{l@{$\quad$}l}
        \ds {1\over2\beta} \left(R\over R_0\right)^{2\beta} 
						& {\rm for\;}\beta\neq0\\[2ex]
        \ds \ln\left(R\over R_0\right)          & {\rm for\;}\beta=0.
\eea \right. \eeq
Here, $R_0$ and $v_0$ are scale radius and scale velocity, and shall be set
to unity in the remainder of this Appendix. The power-law index is restricted
to $\beta\,{\in}\,[{-}\half,\,1]$ with $\beta\eq1$ and ${-}\half$
corresponding to, respectively, the harmonic potential and that created by
a point mass at the origin. The fundamental quantities of circular orbits
expressed as functions of radius, angular momentum, or energy are listed in
Table~\ref{tab:pl}, while the relations for $\kappa$ follow from 
\beq\label{gam-pl}
        \gamma = \sqrt{2/(1+\beta)}.
\eeq

\ifpreprint
  \begin{table}[t]
  \small
  \caption[]{\small
	Relations for circular orbits in power-law potentials \eqb{Phi} in 
	units that imply $R_0\id1\id v_0$. \label{tab:pl}}
  \smallskip
  \begin{tabular}{*{4}{l@{$\;\;$\vline$\;\;$}}l}
  \hline\hline & \multicolumn{4}{c}{} \\[-2ex]
         quantity 
         & \multicolumn{4}{c}{expressed as function of} \\
	 & $R$ & $L$ & $E,\,\beta\nc\neq0$ & $E,\,\beta\eq0$ \\ 
  \hline & & & & \\[-2ex]
  $v$ & $R^\beta$ & $L^{\beta\over1+\beta}$ & 
 	$\left({2\beta\over1+\beta}E\right)^{1\over2}$ & 1 \\[1.5ex]
  $R$ & $R$ & $L^{1\over1+\beta}$ &
	$\left({2\beta\over1+\beta}E\right)^{1\over2\beta}$ & ${\rm e}^{E-1/2}$ 
        \\[1.5ex]
  $L$ & $R^{1+\beta}$ & $L$ &
	$\left({2\beta\over1+\beta}E\right)^{1+\beta\over2\beta}$ & 
	${\rm e}^{E-1/2}$ \\[1.5ex]
  $E,\,\beta\,{\neq}\,0$ & ${1+\beta\over2\beta}R^{2\beta}$ &
	${1+\beta\over2\beta}L^{2\beta\over1+\beta}$ & $E$ & \\[1.5ex] 
  $E,\,\beta\eq0$ & $\half+\ln R$ & $\half+\ln L$ & & $E$ \\[1.5ex] 
  $\Omega$ & $R^{\beta-1}$ & $L^{\beta-1\over\beta+1}$ &
	$\left({2\beta\over1+\beta}E\right)^{\beta-1\over2\beta}$ & 
	${\rm e}^{1/2-E}$
  \\[1ex] \hline\hline
  \end{tabular}
  \end{table}
\else
  \placetable{tab:pl}
\fi

In general, the radial action $J_R$ cannot be expressed in terms of elementary
functions of $(E,L)$; however, for the purely radial orbits with $L\eq0$, $J_R$
may be given in closed form:
\beq\label{JoLc}
	{J_R(L{=}0)\over L_c(E)} = \left\{ \bea{l@{$\quad$}l}
		\ds {(1+\beta)^{1+\beta\over2\beta}\,
			\Gamma\left({1+\beta\over-2\beta}\right)
			\over\sqrt{2\pi}\,|2\beta|^{3/2}\,
                        \Gamma\left({1-2\beta\over-2\beta}\right)} 
						& {\rm for}\;\beta<0,        \\
		\ds \sqrt{{\rm e}\over2\pi}	& {\rm for}\;\beta=0, \\
		\ds {(1+\beta)^{1+\beta\over2\beta}\,
			\Gamma\left({1\over2\beta}\right)
			\over\sqrt{2\pi}\,(2\beta)^{3/2}\,
			\Gamma\left({1+3\beta\over2\beta}\right)} 
						& {\rm for}\;\beta>0.
\eea \right. \eeq

\ifpreprint	\section{T\fL{he} I\fL{sochrone} A\fL{pproximation}}
		\label{app:iso} \noindent
\else		\section{The Isochrone Approximation} \label{app:iso}
\fi
The approximation for $R(t)$ and $\phi(t)$ resulting from the the isochrone
approximation of \Par{iso} are
\ben
\label{iso-Rt}	R^2(t)	&=& (R_E^2+b^2)\, \big(1-e\cos\eta\big)^2 - b^2,\\
\label{iso-tR}	\tR 	&=& \kappa t = \eta - e\,\sin\eta,		\\
\label{iso-e}	e   	&=& \sqrt{\DL} \big/ \big(x_0^2\,\kappa\big)
		     	 = 	\case{\gamma}{2}(1\mi b^{\pr2})
				\sqrt{1-L^2/L_c^2},			\\
\label{iso-j}	\JR 	&=& x_0^2\,\kappa\bigg[1 - \half
			\sum_\pm\sqrt{(1\pm b^\pr)^2\mi e^2}\,\bigg];	\\
\label{iso-phi} \phi(t) &=& \tp + {e\,\hat\op\over\kappa}\sin\eta
			+ {L\over x_0^2\kappa}\sum_{\pm} {A\!\left(\!
			\sqrt{1\pl e\nc\pm b^\pr\over1\mi e\nc\pm b^\pr},
			{\eta\over2}\!\right)
			\over\sqrt{(1\pm b^\pr)^2\mi e^2}},		\\
\label{iso-tp}  \tp	&=& \op t,					\\
\label{iso-op}  \op	&=& \hat\op + {\p g\over\p L},			\\
\label{iso-hop} \hat\op &=& {L\over2x_0^2} 
			\sum_\pm{1\over\sqrt{(1\pm b^\pr)^2\mi e^2}},   \\
\label{iso-g} 	g	&=& \int_{E_c(R_L)}^E\!\!\!\!\!\! \D E\;
			\kappa^2 {\D b^2\over\D E} \left[1\mi {1\over4b^\pr}
			\sum_\pm{\pm(1\nc\pm b^\pr)^3\over
			\sqrt{(1{\pm}b^\pr)^2{-}e^2}}\right]\!
\een
with $b^\pr\id b/x_0$. Here, I have assumed that $b$ was chosen according to 
\eqb{iso-b}, which gives $\D\kappa/\D E\eq{-}3\kappa^{-1}x_0^{-2}$ and
\ben
	L\,x_0^{-2} &=& \Omega(R_E)\sqrt{(1-b^{\pr2}-2e^2\gamma^{-2})
					  (1-b^{\pr2})}			\\
	\kappa^2{\D b^2\over\D E} &=& -10 + {8x_0^4\over3\kappa^2}
			\left[{\D^2\kappa^2\over\D(R^2)^2}\right]_{R_E}.
\een
For the isochrone potential \eqb{henons}, $g\eq0$, while for a flat rotation 
curve,
\beq
	{\p g\over\p L} = -{2v_0^2\over L}\left[1\mi {1\over4b^\pr}
			\sum_\pm{\pm(1\pm b^\pr)^3\over\sqrt{(1{\pm}b^\pr)^2
			{-}e^2}}\right].
\eeq

%
%
\ifpreprint \else
\clearpage
  \begin{deluxetable}{	l@{$\;\;$\vline$\;\;$}
			l@{$\;\;$\vline$\;\;$}
			l@{$\;\;$\vline$\;\;$}
			l@{$\;\;$\vline$\;\;$} l}
  \small
  \tablewidth{0pt}
  \tablecaption{
	Relations for circular orbits in power-law potentials \eqb{Phi} in 
	units that imply $R_0\id1\id v_0$. \label{tab:pl}}
  \tablehead{
         quantity & \multicolumn{4}{c}{expressed as function of} 	\\
	 	  & $R$ & $L$ & $E,\,\beta\nc\neq0$ & $E,\,\beta\eq0$ }
  \startdata
  $v$ & $R^\beta$ & $L^{\beta\over1+\beta}$ & 
 	$\left({2\beta\over1+\beta}E\right)^{1\over2}$ & 1 \\[1.5ex]
  $R$ & $R$ & $L^{1\over1+\beta}$ &
	$\left({2\beta\over1+\beta}E\right)^{1\over2\beta}$ & ${\rm e}^{E-1/2}$ 
        \\[1.5ex]
  $L$ & $R^{1+\beta}$ & $L$ &
	$\left({2\beta\over1+\beta}E\right)^{1+\beta\over2\beta}$ & 
	${\rm e}^{E-1/2}$ \\[1.5ex]
  $E,\,\beta\,{\neq}\,0$ & ${1+\beta\over2\beta}R^{2\beta}$ &
	${1+\beta\over2\beta}L^{2\beta\over1+\beta}$ & $E$ & \\[1.5ex] 
  $E,\,\beta\eq0$ & $\half+\ln R$ & $\half+\ln L$ & & $E$ \\[1.5ex] 
  $\Omega$ & $R^{\beta-1}$ & $L^{\beta-1\over\beta+1}$ &
	$\left({2\beta\over1+\beta}E\right)^{\beta-1\over2\beta}$ & 
	${\rm e}^{1/2-E}$
  \enddata
  \end{deluxetable}
\fi

%
%

\ifpreprint
  \def\thebibliography#1{\section*{R\fL{eferences}}
    \list{\null}{\leftmargin 1.2em\labelwidth0pt\labelsep0pt\itemindent -1.2em
    \itemsep0pt plus 0.1pt
    \parsep0pt plus 0.1pt
    \parskip0pt plus 0.1pt
    \usecounter{enumi}}
    \def\refpar{\relax}
    \def\newblock{\hskip .11em plus .33em minus .07em}
    \sloppy\clubpenalty4000\widowpenalty4000
    \sfcode`\.=1000\relax}
  \def\endthebibliography{\endlist}
\fi

%
%

\ifpreprint \relax \else
\clearpage \onecolumn

\begin{figure}\caption[]{
	Lindblad's classical epicycle theory for the logarithmic potential
	(flat rotation curve). The solid line gives $Y(R)$ \eqb{Y-xi-one}
	while its quadratic approximation \eqb{lb-y} is shown as broken line. 
	The thin horizontal line corresponds to the energy of an orbit with 
	$L\eq0.9L_c(E)$. Note the strong asymmetry of $Y(R)$. \label{fig:lb-y}}
        \end{figure}

\begin{figure}\caption[]{
	Kalnajs' improved epicycle theory for the logarithmic potential.
	The solid line gives $Y(x)$ \eqb{Y-xi-R} while its quadratic 
	approximation \eqb{kl-y} is shown as broken line. The thin horizontal 
	line corresponds to an orbit with $L\eq0.9L_c(E)$. \label{fig:kl-y}}
        \end{figure}

\begin{figure}\caption[]{
	The isochrone approximation for the logarithmic potential. The solid 
	line gives $Y(x)$ \eqb{Y-xi-R} while its quadratic approximation 
	\eqb{iso-y} is shown as broken line. The thin horizontal line 
	corresponds to an orbit with $L\eq0.9L_c(E)$. \label{fig:iso-y}}
        \end{figure}

\begin{figure}\caption[]{
	The Keplerian approximation for the logarithmic potential. The solid 
	line gives $Y(x)$ \eqb{Y-xi-R} while its quadratic approximation 
	\eqb{iso-y} is shown as broken line. The thin horizontal line 
	corresponds to an orbit with $L\eq0.9L_c(E)$. \label{fig:kep-y} }
        \end{figure}

\begin{figure}\caption[]{
	The approximation arising from $\xi\nc\propto R$ and $n\eq\gamma\mi1$ 
	for the logarithmic potential. The solid line gives $Y(x)$ \eqb{Y-xi-R},
	while its quadratic approximation \eqb{crz-y} is shown as broken line.
	The thin horizontal line corresponds to an orbit with $L\eq0.9L_c(E)$.
	\label{fig:crz-y} }
        \end{figure}

\begin{figure}\caption[]{
	The approximations described in \Par{new} for the logarithmic potential.
	The solid line gives $Y(x)$ \eqb{Y-xi-R}, while the dashed and dotted 
	lines correspond to the approximation \eqb{new-y} for $\kmax\eq1$ and
	$\kmax\eq2$, respectively, for the two orbits with $L/L_c(E)\eq0.4$ and
	0.9 (for the latter the approximations can hardly be distinguished from
	the exact relation).
	\label{fig:new-y} }
        \end{figure}

\begin{figure}\caption[]{
	Surfaces of the $R$-$\pR$ section of phase space: comparison of exact 
	orbits in the logarithmic potential, which supports a flat rotation 
	curve, with the predictions of the various orbit approximation of 
	\Par{cases} and \Par{new}. The orbital angular momenta are, from inside
	out, 0.98, 0.95, 0.9, 0.85, 0.8, 0.7, 0.6, and 0.5 of $L_c(E)$.
	\label{fig:RpR} }
        \end{figure}

\begin{figure}\caption[]{
	Surfaces of the $R$-$(\phi\mi\op t)$ section of phase space: 
	comparison of exact orbits in the logarithmic potential, which supports 
	a flat rotation curve, with the approximations for the azimuthal motion.
	The orbits have the same angular momenta as in \Fig{RpR}. Approximations
	which allow for a canonical map to a set of action-angle variables are 
	indicated as `AAV'. Note that multiplying the $y$-axis with $R$, these 
	essentially are the epicycles which the orbits perform around their 
	guiding centers.
	\label{fig:Rphi} }
        \end{figure}

\begin{figure}\caption[]{
	Radial action $\JR$ (left column) and frequency $\oR$ (middle column)
	and azimuthal frequency $\op$ (right column): comparison with exact 
	orbits for the potentials supporting the rotation curves $v_c\nc\propto 
	R^\beta$ with $\beta\eq{-}0.2$, 0, and 0.2 (from top to bottom). Note
	that $\oR\eq\kappa(E)$ for all these approximations except Lindblad's 
	epicycle theory and the approximation of \Par{new} with $\kmax\eq2$
	(\eqj{new-oR}, second order). \label{fig:joo} }
        \end{figure}

%
%

	\clearpage
	\centerline{\epsfxsize=120mm \epsfbox[20 402 580 710]{Dehnen1.fig1.ps}}
        \centerline{Fig.~\ref{fig:lb-y}}

	\vspace{2cm}
	\centerline{\epsfxsize=120mm \epsfbox[20 402 580 710]{Dehnen1.fig2.ps}}
        \centerline{Fig.~\ref{fig:kl-y}}

	\clearpage
	\centerline{\epsfxsize=120mm \epsfbox[20 402 580 710]{Dehnen1.fig3.ps}}
        \centerline{Fig.~\ref{fig:iso-y}}

	\vspace{2cm}
	\centerline{\epsfxsize=120mm \epsfbox[20 402 580 710]{Dehnen1.fig4.ps}}
        \centerline{Fig.~\ref{fig:kep-y}}

	\clearpage
	\centerline{\epsfxsize=120mm \epsfbox[20 402 580 710]{Dehnen1.fig5.ps}}
        \centerline{Fig.~\ref{fig:crz-y}}

	\vspace{2cm}
	\centerline{\epsfxsize=120mm \epsfbox[20 402 580 710]{Dehnen1.fig6.ps}}
        \centerline{Fig.~\ref{fig:new-y}}

	\clearpage
	\centerline{\epsfxsize=155mm \epsfbox[18 324 589 707]{Dehnen1.fig7.ps}}
        \centerline{Fig.~\ref{fig:RpR}}

	\clearpage
	\centerline{\epsfxsize=155mm \epsfbox[18 324 589 707]{Dehnen1.fig8.ps}}
        \centerline{Fig.~\ref{fig:Rphi}}

	\clearpage
	\centerline{\epsfxsize=170mm \epsfbox[21 290 586 710]{Dehnen1.fig9.ps}}
        \centerline{Fig.~\ref{fig:joo}}

\fi 
\end{document}